\documentclass[a4paper,twocolumn,pra]{revtex4-1}
\usepackage{amsmath}
\usepackage{amssymb}
\usepackage{graphicx,color}
\usepackage{amsthm}
\usepackage{hyperref}
\usepackage{extarrows}
\usepackage{adjustbox,lipsum}
\usepackage{makecell}
\usepackage{bbold}

\begin{document}

\newcommand*{\cl}[1]{{\mathcal{#1}}}
\newcommand*{\bb}[1]{{\mathbb{#1}}}
\newcommand{\ket}[1]{|#1\rangle}
\newcommand{\bra}[1]{\langle#1|}
\newcommand{\inn}[2]{\langle#1|#2\rangle}
\newcommand{\proj}[2]{| #1 \rangle\!\langle #2 |}
\newcommand*{\tn}[1]{{\textnormal{#1}}}
\newcommand*{\1}{{\mathbb{1}}}
\newcommand{\T}{\mbox{$\textnormal{Tr}$}}
\newcommand*{\todo}[1]{\textcolor[rgb]{0.99,0.1,0.3}{#1}}

\theoremstyle{plain}
\newtheorem{prop}{Proposition}
\newtheorem{proposition}{Proposition}
\newtheorem{theorem}{Theorem}
\newtheorem{lemma}[theorem]{Lemma}
\newtheorem{remark}{Remark}

\theoremstyle{definition}
\newtheorem{definition}{Definition}

\title{Sample-size-reduction of quantum states for the noisy linear problem}
\author{Kabgyun Jeong}
\email{kgjeong6@snu.ac.kr}
\affiliation{Research Institute of Mathematics, Seoul National University, Seoul 08826, Korea}
\affiliation{School of Computational Sciences, Korea Institute for Advanced Study, Seoul 02455, Korea}

\date{\today}

\begin{abstract}
Quantum supremacy poses that a realistic quantum computer can perform a calculation that classical computers cannot  in any reasonable amount of time. It has become a topic of significant research interest since the birth of the field, and it is intrinsically based on the efficient construction of quantum algorithms. It has been shown that there exists an expeditious way to solve the noisy linear (or learning with errors) problems in quantum machine learning theory via a well-posed quantum sampling over pure quantum states. In this paper, we propose an advanced method to reduce the sample size in the noisy linear structure, through a technique of randomizing quantum states, namely, $\varepsilon$-random technique. Particularly, we show that it is possible to reduce a quantum sample size in a quantum random access memory (QRAM) to the linearithmic order, in terms of the dimensions of the input-data. Thus, we achieve a shorter run-time for the noisy linear problem.
\end{abstract}

\maketitle

%%%%%
\section{\label{sec:Introduction}Introduction}
The Shor's factoring algorithm~\cite{S99}, first proposed in 1994, states that there is an efficient way to find a prime factor $x$ for a given integer $N=a\cdot x$ via quantum Fourier transform (QFT). It changes the computing paradigm from the classical to the quantum regime. Despite being a pivotal method across computational science, there exists a difficulty when a noise is involved. In 2005, Regev introduced a powerful conjecture~\cite{R05} for defense against quantum attacks, known as the learning with errors (LWE) structure, by adding a small-sized Gaussian noise $\delta$ to the factoring problem. The non-trivial problem is expressed as $N'=a\cdot x+\delta$, and we call this task as the \emph{noisy linear problem} (NLP) in a broad sense. This paved the way for post-quantum cryptography to emerge as a modern method of secure communications. While the most quantum algorithms including Shor's one mainly focus on the data-processing with a well-posed quantum sample underlying a superposition principle, however it is also possible to take into account updating the \emph{initial} quantum data (in a feature space), before the quantum algorithm run, for algorithmic efficiency. The quantum data in a higher dimension are essentially different from the shape of classical big-data, i.e., geometrically, any quantum states form a unit hypersphere. Here, we try to reveal the usefulness of initial quantum-data structure in quantum learning theory, and apply it to the specific noisy linear problem. Actually, famous Shannon's noisy channel coding theorem predicts a communication is always robust against a small environmental-noise~\cite{S48}, and it is naturally believed even in the quantum regime.

The noisy linear problem is fundamental in the theory of computer science and machine learning. Particularly, modern cryptography is based on the difficulty of the noisy linear problem, and it was widely believed that it is the strongest candidate for post-quantum cryptography. However, Grilo, Kerenidis, and Zijlstra have recently proposed an efficient quantum algorithm, which is purely based on a quantum oracle for (partially) solving the noisy linear (especially, LWE) scheme through an assumption of the well-posed construction of quantum sample~\cite{GKZ19} (including a binary strategy~\cite{CSS15}), and we call it shortly `GKZ algorithm'. Specifically, if we assume that the quantum sample can be prepared in a quantum superposed state relating to the problem, a kind of quantum Fourier transforms (QFTs)~\cite{S99}, called the Bernstein-Vazirani (BV) algorithm~\cite{BV97} can solve the noisy linear problem within polynomial resources and running times. However, this kind of algorithms has a precondition such that there exists an efficient quantum memory (i.e., QRAM). As pointed before by Aaronson~\cite{A15}, the QRAM issue is not a simple problem so far, but it seems to be a technical one.

%\bigskip
In addition to post-quantum cryptography, quantum-key-distribution (QKD) protocols~\cite{BB84,E91,B92} form another basis for secure quantum communications. As a special case of quantum cryptographic primitives, a random unitary channel (RUC), also known as a randomizing map or a private quantum channel~\cite{AMTW00}, including its discrete~\cite{JK15} and Gaussian variants~\cite{B05,JKL15,BSZ20} in quantum cryptographic schemes, has many crucial implications. It has perfect information-theoretic security, and can be constructed in optimal ways~\cite{NK06,BZ07}. Furthermore, RUC can be directly utilized for an existence-proof for the superadditivity of the classical capacity~\cite{H09,HW08} through a pair of RUCs. Also, it can be utilized for superdense coding scheme~\cite{HHL04} and quantum data-hiding task~\cite{HLSW04}. Actually, RUC is an inverse process of QKD protocols in that two legitimate users with secret keys can always generate the maximally mixed state, and it can be efficiently constructed by quantum randomization techniques. However, we cannot find any studies on quantum algorithms for its speed-up utilizing the quantum randomization techniques. In this paper, we first suggest that quantum randomizing methods (inspired by the efficient RUC-construction) can be directly applied to reduce the initial quantum sample size used in quantum learning theory (i.e., NLP), since the efficiency of the quantum randomizing technique over any quantum states has been analyzed previously~\cite{HLSW04}.

The quantum sample proposed by Grilo \emph{et al.} in the noisy linear problem~\cite{GKZ19} is formally given by
\begin{equation} \label{eq:qs}
\ket{\psi}_{DA}=\frac{1}{\sqrt{q^d}}\sum_{\mathbf{a}\in\bb{F}_q^d}\ket{\mathbf{a}}_D\ket{\mathbf{a}\cdot\mathbf{x}+\delta_{\mathbf{a}}\;\tn{mod}~q}_A\in\cl{P}(\bb{C}^{d+1}),
\end{equation}
where $\mathbf{x}\in\bb{F}_q^d$ is fixed, $\mathbf{a}\in\bb{F}_q^d$ is chosen uniformly at random, and the noise term $\delta_{\mathbf{a}}\in\bb{F}_q$ is an independent and identically distributed (IID) random variable with a certain error probability distribution~\cite{BV14}. Here, $\bb{F}_q$ denotes a finite field of order $q$. Specifically, $\mathbf{a}$ and $\mathbf{x}$ are decomposed into $a_1a_2\cdots a_d$ and $x_1x_2\cdots x_d$, respectively, and $a_j,x_j\in\bb{F}_q$ for any $j$. Our main goal is to recover $\mathbf{x}$ efficiently in this quantum setting. It can be seen that $\cl{P}(\bb{C}^d)$ denotes the set of all \emph{pure} quantum states---a set of unit vectors on the Hilbert space $\bb{C}^d$, and $D$ and $A$ denote the data and ancilla systems, respectively. This process can be prepared by a quantum oracle function $\cl{R}_\psi$ under a help of QRAM, mapping a set of input pure states $\left\{\ket{a_j}_D\right\}_{j=1}^{d}$ and the ancilla $\ket{\mu}_A$ to the output $\ket{\psi}_{DA}$ for every $\mu\in\bb{F}_q$. For a given quantum sample $\ket{\psi}_{DA}$ as in the superposition of pure quantum states, applying QFT (i.e., $\tn{QFT}_q\ket{a}=\frac{1}{\sqrt{q}}\sum_{b=0}^{q-1}\omega^{ab}\ket{b}$ with $\omega=e^{\frac{2\pi i}{q}}$ the root of unity) on the sample state, i.e., $\tn{QFT}_q^{\otimes(d+1)}\ket{\psi}_{DA}$, returns the appropriate output value $\mathbf{x}$ with a high probability. This is known as the Bernstein-Vazirani (BV) algorithm.

%\bigskip
In fact, we wish to find $\mathbf{x}$ efficiently in this quantum learning scenario. In this situation, we may apply the quantum randomizing techniques on $\left\{\ket{a_j}_D\right\}_{j=1}^{d}\subset\cl{P}(\bb{C}^d)$ (via the $\varepsilon$-net construction) over the data systems. Thus, as a new data-set, we can obtain a smaller set of net points $\left\{\ket{\tilde{a_j}}_D\right\}_{j=1}^{m^*}$ ($m^*<d$), where $m^*:=\sqrt{O(d\log d)}$ and $m$ is sufficiently large. We notice that $\sqrt{O(d\log d)}:=O(\sqrt{d\log d})$. This means that we can possibly solve the noisy linear problem more efficiently compared to previous quantum approach~\cite{GKZ19} in the framework of its sample size and running-time.

%\bigskip
Here, we provide that in principle we can reduce the quantum sample-size given by $d$ to $m^*$ via quantum randomization techniques through a mathematical tool known as the $\varepsilon$-net construction~\cite{HLSW04} and L\'{e}vy's inequality~\cite{L51}. These methods discretize a set of pure quantum states to a finite less-sized set on a unit hypersphere, and estimate large deviations for random variables, respectively. (See the technical lemmas in the \textbf{Appendix A} for details.)

%\bigskip
This paper is organized as follows. In Sec.~\ref{grilo}, we briefly introduce the original GKZ algorithm for solving the noisy linear problem in which they make use of well-posed quantum superposed samples. In Sec.~\ref{results}, we propose our main result for reducing the quantum sample-size via quantum randomization techniques. This implies that we can solve NLP more efficiently than GKZ algorithm in the linearithmic order. We also suggest a new type of model for QRAM, namely, approximate QRAM and analyze its performance in Sec.~\ref{aqram}. Finally, discussions and remarks are offered in Sec.~\ref{discussion}, and some open questions are raised for future works.

%%%%%%%%===============
\begin{figure*}%[tbhp]
\centering
\includegraphics[width=0.7\linewidth]{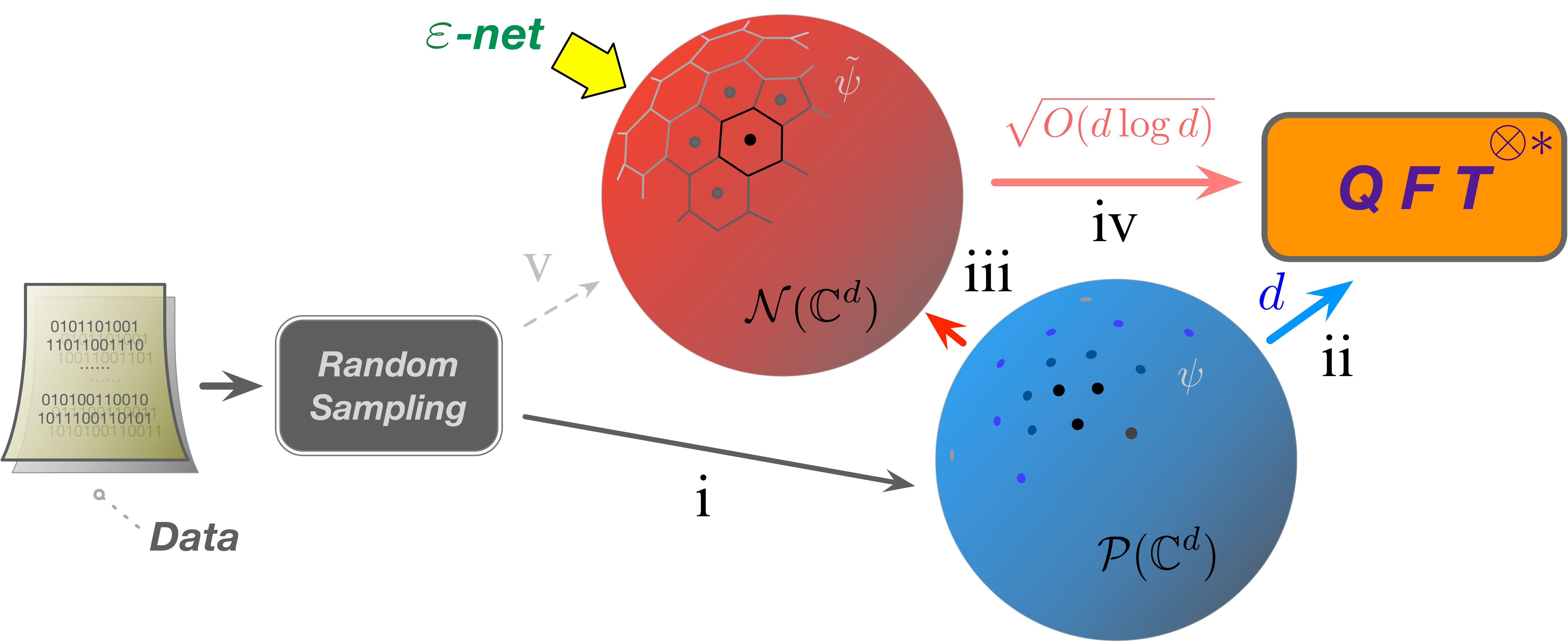}
\caption{\textbf{A model for the quantum state-preparation in an approximate quantum random access memory} (i.e., $\varepsilon$-QRAM): The classical datasets are prepared in sampled \emph{pure} quantum states $\psi\in\cl{P}(\bb{C}^d)$ through a random sampling at step `i'. By using quantum $\varepsilon$-random technique, we can obtain net points $\tilde{\psi}$'s on the unit-sphere at step `iii'. Note that the cardinality of the set of $\varepsilon$-net states (i.e., $m^*=\sqrt{O(d\log d)}$) is less than that of the previously sampled pure states (i.e., $d$). While the Grilo \emph{et. al.}'s quantum algorithm takes a step from i$\to$ii, our algorithm makes use of i$\to$iii$\to$iv before the BV algorithm over QFTs. Alternatively, we expect that the step v$\to$iv is also possible in the QRAM process with a quantum oracle.}
\label{Fig1}
\end{figure*}
%%%%%%%%=============== 
\bigskip

%%%%%
\section{Original GKZ algorithm} \label{grilo}
Before the main result, we shortly review the GKZ algorithm for solving the noisy linear problem. We assume that the errors satisfy $\delta_{\mathbf{a}}=\mathbf{a}\cdot\vec{\eta}$, for all $\vec{\eta}=(\eta_0,\ldots,\eta_{d-1})^T\in\bb{F}_q^d$. The error vector $\vec{\eta}$ is chosen from a certain distribution $\chi$ over $\bb{F}_q$, which is defined by a discrete Gaussian with a small standard deviation at zero. For a given quantum sample in Eq.~(\ref{eq:qs}), the QFTs return the following outcome:
\begin{widetext}
\begin{align}
\left(\tn{QFT}_q^{\otimes d}\otimes\tn{QFT}_q\right)\ket{\psi}_{DA}
=\frac{1}{q^d\sqrt{q}}\sum_{\mathbf{a}\in\bb{F}_q^d}\sum_{\mathbf{b}\in\bb{F}_q^d}\sum_{c\in\bb{F}_q}\omega^{\mathbf{a}\cdot(\mathbf{b}+c\mathbf{x}')}\ket{\mathbf{b}}_D\otimes\ket{c\;\tn{mod}~q}_A,
\end{align}
\end{widetext}
where $\mathbf{x}'=(\mathbf{x}+\vec{\eta})$. By exploiting the delta function, $\Delta_{b_j,-cx_j}:=\tfrac{1}{q}\sum_{a_j\in\bb{F}_q}\omega^{a_j(b_j+cx_j)}$ for each register $j\in\{0,\ldots,d-1\}$, we can obtain
\begin{equation}
\frac{1}{\sqrt{q}}\sum_{\mathbf{b}\in\bb{F}_q^d}\sum_{c\in\bb{F}_q}\ket{-c\mathbf{x}'}_D\otimes\ket{c\;\tn{mod}~q}_A.
\end{equation}
However, $\mathbf{x}'$ is not equal to the true value $\mathbf{x}$, when $\vec{\eta}\neq0$ (without the noise term, i.e., $\vec{\eta}=0$, we can directly recover the hidden $\mathbf{x}$ by simply measuring the ancilla register $A$). Thus, we need to check if $\mathbf{x}'=\mathbf{x}$ is satisfied by substituting $\mathbf{b}=-c\mathbf{x}$, formally called the `test' process, and by calculating the success probability as
\begin{align}
\tn{Pr}\left[\mathbf{x}'=\mathbf{x}\right]&=\frac{1}{q^{2d+1}}\left\|\sum_{c\in\bb{F}_q}\sum_{\mathbf{a}\in\bb{F}_q^d}\omega^{c\delta_{\mathbf{a}}}\ket{-c\mathbf{x}}_D\otimes\ket{c}_A\right\|^2 \nonumber\\
&\ge\frac{\alpha}{t}\cos^2(2\pi\alpha),
\end{align}
where $\alpha\in[0,1/4)$ and $c\le\tfrac{\alpha q}{t}$. Furthermore, we notice that the condition $|\mathbf{a}'\cdot\mathbf{x}+\delta_{\mathbf{a}'}-\mathbf{a}'\cdot\mathbf{x}'|\le t$ needs to be satisfied for every randomly chosen test sample $\mathbf{a}'\in\bb{F}_q^d$. Under the condition over $M$ trials, we complete the test accepting $\mathbf{x}'=\mathbf{x}$, that is, the fail probability is given by $\tn{Pr}\left[\mathbf{x}'\neq\mathbf{x}\right]\le\left(\tfrac{2t+1}{q}\right)^M$ (see details the Lemma 1 in Ref.~\cite{GKZ19}). This statement will be also updated for our approximate scheme.

%\bigskip
In Refs.~\cite{A15,GKZ19}, as a hard problem it was seriously pointed out that a largely superposed state must be prepared in the sampling process---This issue is generally called a QRAM problem~\cite{GLM08,KP17} (see also Ref.~\cite{E21}). Here, our approach for the noisy linear solving quantum algorithm aims to construct an approximate QRAM (or $\varepsilon$-QRAM) through an $\varepsilon$-net analysis as in Fig.~\ref{Fig1}, when the initial quantum samples are all in pure quantum states. We discuss $\varepsilon$-QRAM thoroughly later in this paper.

%%%%%
\section{Main Result} \label{results}
We present the details of the quantum randomizing methods and their performance to solve the noisy linear problem. As mentioned before, we should remember that all communications are not so susceptible to a noise by the Shannon's noisy channel coding theorem. 

First of all, let us recapitulate the mathematical notations. Let $\cl{B}(\bb{C}^d)$ be the space of (bounded) linear operators on the $d$-dimensional Hilbert space $\bb{C}^d$, and $\cl{U}(d)\subset\cl{B}(\bb{C}^d)$ be the unitary group on the Hilbert space. Note that $\1_d$ is simply the $d\times d$ identity matrix on the space. For simplicity, we denote a pure state $\psi:=\proj{\psi}{\psi}\in\cl{P}(\bb{C}^d)$. In addition, we assume that a quantum channel $\Lambda$ is a completely positive and trace-preserving (CPT) map. While we make use of $m=d$ for an input quantum sample-size, we denote $m=d^2$ as the number of unitary operations in the case of quantum channel. Now, let us define an $\varepsilon$-randomizing map with respect to the Schatten $p$-norm in the trace class~\cite{J12}, which is an \emph{extended} version of the statement proposed by Hayden \emph{et al}. ~\cite{HLSW04}, and it is defined as follows.

%\bigskip
For every quantum state $\varrho\in\cl{B}(\bb{C}^d)$, we call a CPT map $\Lambda:\cl{B}(\bb{C}^d)\to\cl{B}(\bb{C}^d)$ an \emph{$\varepsilon$-randomizing} map with respect to the Schatten $p$-norm, if 
\begin{equation}
\left\|\Lambda(\varrho)-\frac{\1_d}{d}\right\|_p\le\frac{\varepsilon}{\sqrt[p]{d^{p-1}}},
\end{equation}
where $\1_d/d$ denotes the $d$-dimensional maximally mixed state (MMS), and $\|M\|_p:=\left(\sum_{j=1}^d|s_j|^p\right)^{1/p}=(\T(M^\dag M)^{p/2})^{1/p}$ is the Schatten $p$-norm for any matrix $M\in\cl{B}(\bb{C}^d)$ ($1\le p\le\infty$) with singular values $\{s_j\}_{j=1}^d$.

%\bigskip
In general, CPT map $\Lambda$ acting on the quantum state $\varrho$ can be naturally constructed by
\begin{equation} \label{pqc}
\Lambda(\varrho)=\frac{1}{m}\sum_{j=1}^mU_j\varrho U_j^\dag,
\end{equation}
where the unitary operator $U_j$'s are chosen uniformly at random from the unitarily invariant measure (or Haar measure) on the unitary group $\cl{U}(d)$, and $m$ is the number of unitary operators depending on the input dimension $d$. It is known that $m=d^2$ is optimal to create a perfect $d$-dimensional maximally mixed state. However, we propose that $m=O(d\log d)$ is also sufficient to create MMS in the limit of $d\to\infty$~\cite{HLSW04,DN06,J12}. This reduction will be exploited for constructing an efficient quantum samples used in the noisy linear problem, and thus we call it as `\emph{$\varepsilon$-random technique}' for quantum algorithms. Here, we notice that it is important to choose $m$ sufficiently large.

%\bigskip
It is also worth noting that if $d\mu$ is the unitarily invariant measure chosen uniformly at random on the unitary group $\cl{U}(d)$, then, for every quantum state $\varrho\in\cl{B}(\bb{C}^d)$, it satisfies
\begin{equation} \label{invmeas}
\int_{\cl{U}(d)}U\varrho U^\dag d\mu=\frac{\1_d}{d},\;\;\;\;\forall U\in\cl{U}(d).
\end{equation}

For example, when $d=2$, the above unitaries in the identity of Eq.~(\ref{pqc}) take the form of a set of Pauli matrices including $\1_2$.
The fundamental feature of the unitarily invariant measure in Eq.~\eqref{invmeas} is that it assures of the uniform randomness for given any type of quantum states. If the states are given in the form of a quantum sample, in principle the measure can be used to check that the sample is uniform or not. Because most algorithms essentially require a random sample as its input in the regime of the classical as well as quantum. This observation offers the possibility for us to construct the efficient private quantum channel $\Lambda$ from $m=d^2$ onto $m=O(d\log d)$ unitary operations under a condition of $m\gg1$. (See \textbf{Appendix B} for the detailed proof of {\bf Theorem 1} below.) Consequently, it is sufficient to create RUC (or $\varepsilon$-randomizing map) in Eq.~\eqref{pqc} with a randomly chosen unitary operations of $O(d\log d)$. Our strategy is based on this reduction technique, and applies it to the regime of the quantum sampling problem, and the full statement of $\varepsilon$-randomizing map is given as follows:

\bigskip
\textbf{Theorem 1}. 
Let $\varepsilon>0$ and the dimension $d$ be sufficiently large. There exists a set of unitary operators $\{U_j\}_{j=1}^m\subset\cl{U}(d)$ with cardinality 
\begin{equation} \label{rucp}
m\ge\frac{\kappa}{\varepsilon^2}d\log\left(\frac{10d^{(p-1)/p}}{\varepsilon}\right)
\end{equation}
such that $\Lambda(\varphi)=\frac{1}{m}\sum_{j=1}^mU_j\varphi U_j^\dag$ on $\cl{B}(\bb{C}^d)$ is $\varepsilon$-randomizing map with respect to the Schatten $p$-norm (for any $p\ge1$), and $\kappa$ is a universal constant. 

\bigskip
For convenience, we exploit the notation $\tilde{m}:=O(d\log d)(={m^*}^2)$ instead of $m$ as an upper bound. From {\bf Theorem 1} in the field of quantum channel theory, we try to highlight on the quantum sample-reduction problem over quantum algorithms, especially, relating to the solvability of the noisy linear structure in quantum machine learnings. Here, we further improve the GKZ algorithm in the order of linearithmic over a superposed quantum sample. Here, we notice that we will take an upper bound (i.e., $O(d\log d)$) rather than the exact lower bound in Eq.~(\ref{rucp}). Before presenting the main result, let us observe how quantum randomizing technique works. The intuition is basically conducted by following two lemmas. The first one is the `$\varepsilon$-net' construction, and the second one is `L\'{e}vy's inequality' on the set of pure quantum states. We call, such a combination of two lemmas, as $\varepsilon$-random technique as mentioned before.

\bigskip
\textbf{Lemma 2} ($\varepsilon$-net~\cite{HLSW04}).
Let $\varepsilon>0$, and the dimension $d$ be sufficiently large. For any pure quantum state $\ket{\psi}\in\cl{P}(\bb{C}^d)$, we can choose a net-point $\ket{\tilde{\psi}}\in\cl{N}(\bb{C}^d)\subset\cl{P}(\bb{C}^d)$ satisfying $\|\psi-\tilde{\psi}\|_1\le\varepsilon$, where $\|\cdot\|_1$ is the trace norm. Then, there exists a set $\cl{N}(\bb{C}^d)$ of pure states such that
\begin{equation} \label{e-net}
|\cl{N}(\bb{C}^d)|\le\left(\frac{5}{\varepsilon}\right)^{2d},
\end{equation}
where $|\cdot|$ denotes the cardinality of the net $\cl{N}$. Also, recall that $\psi:=\proj{\psi}{\psi}$.

\bigskip
\textbf{Lemma 3} (L\'{e}vy's inequality~\cite{L51}).
Let $F$ be a function $F:\partial\cl{S}^d\to\bb{R}$ defined on the $d$-dimensional \emph{unit} ball $\cl{S}^d$ and its boundary $\partial\cl{S}^d$. Suppose that a point $\psi\in\partial\cl{S}^d$ is chosen uniformly at random. Then, for every $\varepsilon'>0$,
\begin{equation} \label{Levy}
\tn{Pr}[|F(\psi)-\bb{E}(F)|\ge \varepsilon']\le C_1\exp\left(-\frac{C_2d\varepsilon'^2}{\gamma^2}\right),
\end{equation}
where $\gamma:=\sup|\nabla F|$ is the Lipschitz constant of $F$, and $C_1$ and $C_2$ are universal constants.

\bigskip
Now, let us define an abstract quantum channel $\tilde{\Lambda}$, which convert a set of pure quantum states $\{\psi\}_{j=1}^m$ to another less-sized set of pure quantum states $\{\tilde{\psi}\}_{j=1}^{m^*}$ on the net space by exploiting {\bf Lemma 2} and {\bf Lemma 3}. Here, we exploit $m$ as in the notion of quantum sample-size. This virtual process $\tilde{\Lambda}$ can be performed by $\varepsilon$-QRAM (see Subsec.~\ref{aqram}). Now, we fixed $m=d$ and $m^*=\sqrt{O(d\log d)}$ for accordance between RUC ($\Lambda$) and $\varepsilon$-QRAM ($\tilde{\Lambda}$): it changes the representation from density operator to state one. We notice that, in the framework of the quantum sample preparation of the GKZ algorithm, when the initial sample may start with $d\to\infty$, then the success probability of the algorithm can be more increased with high probability. The Fig.~\ref{Fig1} conceptually shows a quantum state-preparation for the NLP-solving sample in the $\varepsilon$-QRAM subroutine.

\bigskip
By using those ingredients with $\varepsilon$-QRAM, we are ready to suggest our main result as follows, and the proof is essentially equivalent to the GKZ algorithm~\cite{GKZ19}, where Eq.~\eqref{eq:qs} on the original quantum-sample is substituted as a netized sample. For convenience, we also make use of the terms such as `Test Candidate ({\bf TC})' and `NLP Algorithm ({\bf NLP-A})' defined in Ref.~\cite{GKZ19} for the proof.

\bigskip
\textbf{Proposition 4} (Main result).
Let $d$ be sufficiently large (thus, $m^*$ is) and $q$ be a prime number. Assuming we can efficiently prepare a superposed quantum sample through the $\varepsilon$-QRAM over $\cl{N}(\bb{C}^d)\subset\cl{P}(\bb{C}^d)$ in the form of
\begin{equation} \label{eq:netqs}
|{\tilde{\psi}}\rangle_{DA}=\frac{1}{\sqrt{q^{m^*}}}\sum_{\tilde{\mathbf{a}}\in \bb{F}_q^{m^*}}\ket{\tilde{\mathbf{a}}}_D\ket{\tilde{\mathbf{a}}\cdot \tilde{\mathbf{x}}+\delta_{\tilde{\mathbf{a}}}\;\tn{mod}~q}_A,
\end{equation}
where $\delta_{\tilde{\mathbf{a}}}$ is a random variable chosen noise distribution with maximum noise magnitude $t=poly(m^*)$. Then, there \emph{exists} a quantum algorithm that outputs $\tilde{\mathbf{x}}$ (such that $|\tilde{\mathbf{x}}|<|\mathbf{x}|$) with probability $\frac{1}{20tq^{m^*-1}}$.
\bigskip

{\it Proof}. Let $L$ and $M$ be parameters in order to prove the proposition. By using {\bf Lemma 8} (which is the modification of Lemma 1 in Ref.~\cite{GKZ19}) in Appendix C, i.e., formally, the {\bf TC} accepts with probability
\begin{align}
\tn{\bf TC}~(\mathbf{x}',M)
&=\left\{ \begin{array}{ll} 
1,\ & \;\tn{if}\;\;\mathbf{x}'=\tilde{\mathbf{x}};\\ & \\
\le(\tfrac{2t+1}{q})^M,\ & \; \tn{if}\;\;\mathbf{x}'\neq\tilde{\mathbf{x}}.
\end{array}\right.
\end{align}
By using the union bound, the probability that at least one of independent $L$-call to $\tn{\bf TC}~(\mathbf{x}',\log\eta^{-1})$ accepts some $\mathbf{x}'\neq\tilde{\mathbf{x}}$ is at most $\left({3t}/{q}\right)^M L$. Also from {\bf Lemma 9} (which is the modification of Theorem 1 in Ref.~\cite{GKZ19}), the probability that $\mathbf{x}$ is not the output of independent $L$-call to BV algorithm is at most $\left(1-\frac{\ell}{20tq^{m^*}}\right)^L$.

%%%%%%%%===============
\begin{figure*}
\centering
\includegraphics[width=0.95\linewidth]{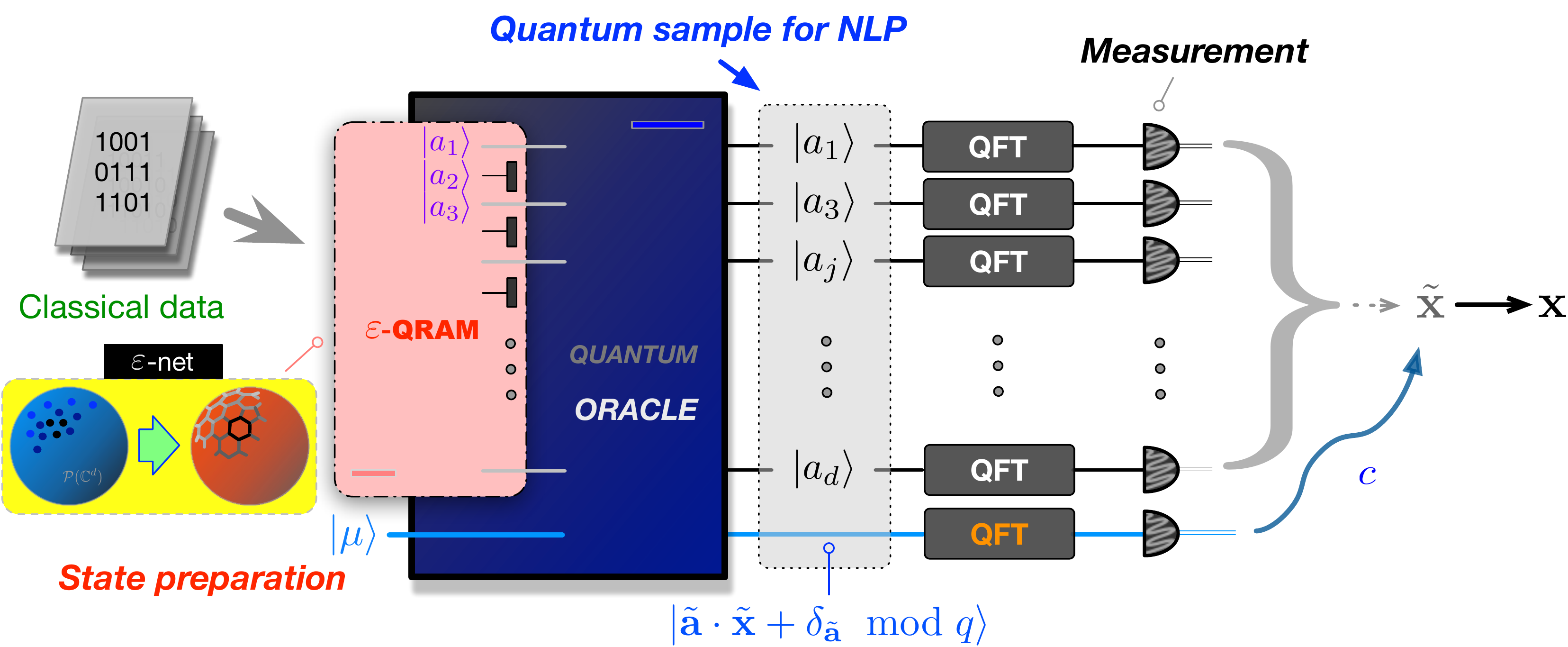}
\caption{\textbf{A schematic diagram of procedure in the NLP-solving quantum algorithm and the abstract notion of $\varepsilon$-QRAM}. In the state preparation step, $\varepsilon$-QRAM ($\tilde{\Lambda}$) transforms pure quantum data into a netized sample through specific blocking operations ($-\!|\!\!|\!\!|\!\!|\!\!|\!\!|\!\!|$~), i.e., the output quantum data are given in the form $\ket{\psi}_{DA}=\ket{\tilde{\mathbf{a}}}_D\otimes\ket{\mu}_A$. Here, we assume that a quantum oracle $\cl{R}_\psi$ via $\varepsilon$-QRAM operates a specific hidden transformation on the input quantum data, and then prepares a quantum sample $|\tilde{\psi}\rangle_{DA}$ as in Eq.~\eqref{eq:netqs}, to perform the desired noisy linear task. However, our quantum sample-size is linearithmically reduced compared to the original quantum sample $\ket{\psi}_{DA}$ in Eq.~\eqref{eq:qs}. According to the BV kernel method (i.e., $\tn{QFT}_q^{\otimes (\lceil m^*\rceil+1)}$ where $m^*=\sqrt{O(d\log d)}$) and following proper quantum measurements, we can efficiently recover the secret information $\mathbf{x}$ hidden in the noisy linear encodings. We notice that $\tilde{\mathbf{x}}$ is $\varepsilon$-close to $\mathbf{x}$ for sufficiently large $d$.}
\label{Fig2}
\end{figure*}
%%%%%%%%===============

By exploiting the union bound again, we address that $\tn{\bf NLP-A}~(L,M)$ does not output $\tilde{\mathbf{x}}$ with probability at most
\begin{equation}
\left(1-\frac{\ell}{20tq^{m^*}}\right)^L+\left(\frac{3t}{q}\right)^M L,
\end{equation}
where we can choose $\ell=q^{m^*}$, $L=20t\log\eta^{-1}$, and $M=1$ for completion of the proof.
$\blacksquare$
\bigskip

This result implies that, for every $\eta>0$, the BV algorithm achieves sample complexity $O(m^*\log\eta^{-1})$ and running-time in $poly(m^*,\log\eta^{-1})$ with probability $1-\eta$ as in the Theorem 2 in Ref.~\cite{GKZ19}. We notice that the set of classical information $\tilde{\mathbf{x}}$ (representing on the net space) is $\varepsilon$-close to the original information of $\mathbf{x}$ from the quantum $\varepsilon$-random technique above.

%\bigskip
We can summarize our total procedure intuitively in terms of $\varepsilon$-QRAM, BV algorithm, and quantum measurement as follows:
\begin{widetext}
\begin{equation}
\ket{\mathbf{a}}_D\xlongrightarrow{}\ket{\psi}_{DA}\xlongrightarrow[\tn{\textbf{Lemma~2,3}}]{\varepsilon\tn{-QRAM}}|\tilde{\psi}\rangle_{DA}\xlongrightarrow{\tn{BV}}\tn{QFT}_q^{\otimes\left\lceil m^*\right\rceil+1}|\tilde{\psi}\rangle_{DA}\xlongrightarrow{\tn{Meas}} \tilde{\mathbf{x}}\sim\mathbf{x}.
\end{equation}
\end{widetext}
The last step denotes a quantum measurement performed on the computational basis, and $\tilde{\mathbf{x}}$ is $\varepsilon$-close to the original secret $\mathbf{x}$. (See Fig.~\ref{Fig2} for details.)

%%%%%%%%%%%%%%
\begin{table*}
\centering
\tabcolsep=0.05in
\begin{adjustbox}{max width=\textwidth}
\begin{tabular}{c c c} 
\hline\hline
{} & \makecell{Quantum algorithms} & \makecell{Best classical algorithms} \\
\hline
\makecell{Integer factoring} 		& $O(\log{d})$~\cite{S99} 		& $2^{o(d)}$ \\
\makecell{Database search} 	& $O(\sqrt{d})$~\cite{G97} 		& $O(d)$ \\
\makecell{HHL} 			& $O(\log{d})$~\cite{HHL09} 		& $O(d)$ \\
\makecell{SVM} 			& $O(\log{d})$~\cite{RML14,HCT+19} 	& $O(\tn{poly}(d))$ \\
\makecell{Noisy linear problem} 	& $O(d)$+${}^\star$QRAM~\cite{GKZ19,CSS15,SLJ+19,SLJ+21}, \textbf{This work} & $2^{O(d)}$~\cite{BKW03}+RAM \\
\makecell{TDA}    & $O(\log^5{d})$~\cite{LGZ16}		& $O(d^2)$ \\
\makecell{Recommendation}    & $O(\tn{poly}\log{d})$~\cite{KP17} 	& ${}^\sharp O(\tn{poly}\log{d})$~\cite{T19,CGL+19} \\
\hline\hline
\end{tabular}
\end{adjustbox}
\bigskip
\caption{\label{tab:QA} \textbf{Comparison of sample-complexity for representative quantum algorithms versus best known classical algorithms}. Here, $d$ is the input dimension in $\bb{F}_q^d$ for the given computational problem. The famous Shor algorithm~\cite{S99} and Grover search algorithm~\cite{G97} can achieve quantum speedups in exponential and quadratic order, respectively. Furthermore, the Harrow-Hassidim-Lloyd (HHL)~\cite{HHL09} sparse matrix inversion, the quantum support-vector-machine (SVM)~\cite{RML14,HCT+19}, and the quantum topological data analysis (TDA)~\cite{LGZ16} algorithm in quantum machine learning theory have remarkable quantum advantages. Beside, for the recommendation system~\cite{KP17}, a classical algorithm can achieve quantum efficiency and it was known as quantum-inspired algorithm~\cite{T19,CGL+19} (${}^\sharp$). For the noisy linear problem (NLP), the GKZ algorithm~\cite{GKZ19} solve the problem in polynomial time, in which there is an assumption that an well-superposed quantum sample (i.e., Eq.~\eqref{eq:qs}) can be efficiently prepared by the quantum random access memory (QRAM denoted by (${}^\star$)). We notice that the analysis of the sample-complexity for the noisy linear model in classical regime is described in Ref.~\cite{BKW03}. In this work, we improve the QRAM-complexity for preparing a quantum sample for NLP by exploiting the quantum $\varepsilon$-random technique, and the subroutine is called as `approximate QRAM (or $\varepsilon$-QRAM)'. Consequently, this work improves the NLP-solving algorithm in the linearithmic order than the GKZ algorithm~\cite{GKZ19} including Ref.~\cite{CSS15,SLJ+19,SLJ+21}.}
\end{table*}
%%%%%%%%%%%%%%

%%%%%
\subsection{Approximate QRAM issue} \label{aqram}
In computer architecture, the random access memory (RAM) plays a crucial role storing and calling out data in computation process. QRAMs~\cite{GLM08,KP17} are quantum analogues of classical RAMs, and they perform similar actions of reading-out and returning datasets in the form of quantum states. However, this case is subtle. The quantum data can be superposed even in the form of entangled state~\cite{A15} because of the following reason: The QRAM performs that
\begin{equation}
\tn{QRAM}:\sum_j\alpha_j\ket{j}\ket{0}\mapsto\sum_j\alpha_j\ket{j}\ket{d_j},
\end{equation}
where $d_j$ is a quantum datapoint corresponding to the memory address $j$, and $\ket{0}$ an ancillary state. Hence, the data set loads in quantum superposition. Assuming that a (quantum) big-data set, the problem could be intriguing for quantum learning theory~\cite{PML14,LMR14,BWP+17}. Unfortunately, this problem also occurs at the proposed approximate QRAMs, but our claim is that it is theoretically improvable, when the given sets are \emph{pure} quantum states. Geometrically, all pure states lie at the boundary on the Bloch sphere. In this case, we can concentrate a quantum information into a net-point on the unit sphere as discussed above, and the role of `$\varepsilon$-QRAM' through the $\varepsilon$-net construction and the Levy's inequality is described in Fig.~\ref{Fig2}.

%\bigskip
The reduction of information resources, starting from the origin of Shannon's data compression theory~\cite{S48}, is a core ingredient in information sciences as well as in quantum computing and quantum simulation~\cite{CMN+18}. While classical information datasets have a diverging hypercubic structure in general, they are intractable. Quantum datasets are geometrically simple in that they have the shape of a unit-hypersphere (see \textbf{Appendix A} for the details of discretization method via $\varepsilon$-net on a higher dimensional unit sphere, i.e., on $d$-dimension pure quantum states). 

%\bigskip
Specifically, the $\varepsilon$-QRAM acts in operational ways to discretize all pure quantum states over the net space in this noisy linear problem, i.e., it outputs a random sample $\ket{\tilde{\mathbf{a}}}$ (instead of the total sample-size of $\ket{\mathbf{a}}$):
\begin{equation}
\sum_j\alpha_j\ket{j}\ket{\mathbf{a}}_D\otimes\ket{\mu}_A\mapsto\sum_j\alpha_j\ket{j}\ket{\tilde{\mathbf{a}}}_D\otimes\ket{\mu}_A,
\end{equation}
which in principle can be used as a quantum sample to resolve the NLP. From this, the quantum oracle gives birth to a quantum sample for solving NLP in terms of $\ket{\tilde{\mathbf{a}}}_D\otimes\ket{\tilde{\mathbf{a}}\cdot\tilde{\mathbf{x}}+\delta_{\mathbf{\tilde{a}}}\;\tn{mod}~q}_A$.  As a comparable method, we can also devise and modify the bucket-brigade QRAM~\cite{GLM08,GLM08a,AGJ+15} in this framework of $\varepsilon$-QRAM structure.

%%%%
\section{Discussion} \label{discussion}
The noisy linear problem is constituted through a classical algorithm, fundamentally adding a small Gaussian noise to the meaningful original dataset. It is generally believed to be intractable in the presence of super-powered quantum computers. However, contrary to popular belief, quantum computing or quantum communication are very powerful, and recent studies showed that it is not true under the assumption of a well-posed quantum sample under a quantum random access memory, by applying quantum Fourier transforms (known as the Bernstein-Vazirani algorithm), as in the case of the Shor's factoring algorithm. Moreover, the difficulty of the noisy linear algorithmic problem is alleviated using the framework of quantum learning theory. In this study, we attempted to improve its efficiency by exploiting the quantum $\varepsilon$-random technique over all pure quantum states. This argument is always possible, if we can freely access a dataset of pure quantum states, because such quantum states have the simple geometric structure of a unit hypersphere. However, there exists an uncomfortable problem to overcome for its efficiency argument, and it is commonly known as QRAM issue. Here, we tried to provide a new type of method represented by $\varepsilon$-QRAM.

More precisely, we proposed an efficient way of a linearithmic reduction for a quantum sample size from $d$ to $m^*=\sqrt{O(d\log d)}$ in the input dimension of the data-set through the quantum $\varepsilon$-random technique, especially, for solving the noisy linear problem. The quantum randomizing techniques over quantum algorithms, inspired by constructing the random unitary channel, rely on purely mathematical tools known as the $\varepsilon$-net and L\'{e}vy's inequality. The key point is that we can always reduce the consumption of the quantum sample before the expensive QFT runs; thus, the number of QFTs can be also reduced in linearithmic order. The impact of this construction in quantum machine learnings is potentially large, for examples, a reduction of quantum Fourier transforms in the circuit-realization as well as in quantum topological data analysis and quantum-inspired algorithms. (See \textbf{Table I}.) A caveat of this is that the non-practical Haar measure problem occurs. However, a specific method such as unitary design~\cite{HHH05,DCEL09} and compressive measurement~\cite{KTA+20} settles the theoretical caveats to be a tractable for selecting a proper set of random unitaries. Furthermore, the cardinality of the unitary set can be further tightened to be $m=O(d)$ with a sharp mathematical argument~\cite{A09}, as long as $d\to\infty$ in random matrix theory.

While quantum advantage can be achieved in quantum algorithms including the noisy linear problem under the assumption of a well-posed quantum sample, it requires access to a largely-superposed quantum state. However, a quantum-inspired (or dequantizing) algorithm recently proposed in Refs.~\cite{T19,CGL+19}, as well as in hybrid algorithms~\cite{H20} could be good candidates to avoid the drawbacks of QRAMs. We also believe that it could be possible to improve via the subtle net analysis (e.g., a mathematical extension of the L\'{e}vy's or McDiarmid's inequalities) on quantum samples, and a divide-and-conquer strategy also can be devised in the quantum learning theory for realizing a noisy intermediate-scale quantum computing. (See also Refs. in \textbf{Table I}.)

\begin{acknowledgments}
This work was supported by the National Research Foundation of Korea (NRF) through a grant funded by the Ministry of Science and ICT (NRF-2020M3E4A1077861, NRF-2022M3H3A1098237) and the Ministry of Education (NRF-2021R1I1A1A01042199).
\end{acknowledgments}

\section*{Data Availability}
The data that support the findings of this study are available within the article.

\section*{Appendix}
Here, we derive the theorem, which inspire the efficient solvability on the noisy linear problem: Let $\varphi:=\proj{\varphi}{\varphi}\in\cl{P}(\bb{C}^d)$ be a pure quantum state and $d\mu$ be the unitarily invariant measure on the unitary group $\cl{U}(d)$. Then, we obtain the $\varepsilon$-randomizing map in the Schatten $p$-norm.

\bigskip
\textbf{Theorem 1}. 
Let $\varepsilon>0$ and the dimension $d$ be sufficiently large. There exists a set of unitary operators $\{U_j\}_{j=1}^m\subset\cl{U}(d)$ with cardinality 
\begin{equation} %\label{rucp}
m\ge\frac{\kappa}{\varepsilon^2}d\log\left(\frac{10d^{(p-1)/p}}{\varepsilon}\right)
\end{equation}
such that $\Lambda(\varphi)=\frac{1}{m}\sum_{j=1}^mU_j\varphi U_j^\dag$ on $\cl{B}(\bb{C}^d)$ satisfies the $\varepsilon$-randomizing map with respect to the Schatten $p$-norm (for any $p\ge1$), and $\kappa$ is a universal constant. Here, we denote the lower bound of $m$ as $\tilde{m}:=O(d\log d)$.

\subsection{Technical lemmas}
We need several technical lemmas for the proof of the result ({\bf Theorem 1}).

\bigskip
\textbf{Lemma 5}.
For any $r,p$ such that $r>p\ge1$, a quantum state $\varrho\in\cl{B}(\bb{C}^d)$ satisfies
\begin{equation}
\left\|\varrho-\frac{\1_d}{d}\right\|_p^r\le d^{\frac{r-p}{p}}\|\varrho\|_r^r-\frac{d^{(r-p)/p}}{d^p},
\label{holder}
\end{equation}
where $\|M\|_p:=\left(\sum_{j=1}^d|s_j|^p\right)^{1/p}=(\T(A^\dag A)^{p/2})^{1/p}$ is the Schatten $p$-norm in the trace class for a matrix $M\in\cl{B}(\bb{C}^d)$ ($1\le p\le\infty$) with singular values $\{s_j\}_{j=1}^d$ of $M$~\cite{B97}.
\bigskip

{\it Proof}. The inequality in Eq.~(18) is straightforward from the fact that $\varrho$ is a density matrix and from the H\"{o}lder's inequality. $\blacksquare$

\bigskip
\textbf{Lemma 6}.
Let $\varphi\in\cl{P}(\bb{C}^d)$ be a fixed pure state. If we define a random variable $Y_{[\varphi]}=\left\|\Lambda(\varphi)-\frac{\1_d}{d}\right\|_p$, then the following inequality holds (for all $r>p\ge1$)
\begin{equation}
\bb{E}_{\{U_j\}}Y_{[\varphi]}\le\left(\frac{\sqrt[p]{d}}{m^p}+\frac{r}{m^{p-1}\cdot\sqrt[p]{d}}\right)^{1/r},
\end{equation}
where the expectation $\bb{E}:=\bb{E}_{\{U_j\}}$ is taken over a set of unitary matrices $\{U_i\}$ chosen at random.
\bigskip

{\it Proof}. (i) $p=1$ and $r=2$ case: We recall that $\Lambda(\varphi)=\frac{1}{m}\sum_{j=1}^mU_j\varphi U_j^\dag$ and $Y_{[\varphi]}=\left\|\Lambda(\varphi)-\frac{\1_d}{d}\right\|_1$. Then, we have
\begin{align} \label{p1r2}
\bb{E}\|\Lambda(\varphi)\|_2^2&=\frac{1}{m}+\frac{1}{m^2}\sum_{j\neq k}^m\bb{E}\T(U_j\varphi U_j^\dag U_k\varphi U_k^\dag) \nonumber\\
&\le\frac{1}{m}+\T\left(\int_{\cl{U}(d)}U_j\varphi U_j^\dag d\mu\cdot\int_{\cl{U}(d)}U_k\varphi U_k^\dag d\mu\right) \\
&=\frac{1}{m}+\T\frac{\1_d}{d^2}=\frac{1}{m}+\frac{1}{d}, \nonumber
\end{align}
where Eq.~(20) comes from the definition of the unitarily invariant measure (i.e., Eq.~(\ref{invmeas})) with IID unitary sets in the index $j,k$. By exploiting the Cauchy-Schwartz inequality, we obtain $Y_{[\varphi]}^2\le d\|\Lambda(\varphi)\|_2^2-1$, and we can obtain $\bb{E}Y_{[\varphi]}^2\le d\bb{E}\|\Lambda(\varphi)\|_2^2-1$~\cite{DN06}. Thus, for a sufficiently large $d$,
\begin{equation}
\bb{E}Y_{[\varphi]}\le\sqrt{\bb{E}Y_{[\varphi]}^2}\le\sqrt{d\bb{E}\left\|\Lambda(\varphi)\right\|_2^2-1}=\sqrt{\frac{d}{m}}.
\end{equation}
(ii) $p=2$ and $r=3$ case: Now, suppose that $Y_{[\varphi]}=\left\|\Lambda(\varphi)-\frac{\1_d}{d}\right\|_2$. As in case (i), we can straightforwardly obtain the inequality: $\bb{E}\|\Lambda(\varphi)\|_3^3\le\frac{1}{m^2}+\frac{3}{md}+\frac{1}{d^2}$. Thus, by using the H\"{o}lder's inequality on the Schatten $p$-norm,
\begin{equation}
\bb{E}\left\|\Lambda(\varphi)-\frac{\1_d}{d}\right\|_2^3\le\sqrt{d}\bb{E}\left\|\Lambda(\varphi)\right\|_3^3-d^{-3/2}\le\frac{\sqrt{d}}{m^2}+\frac{3}{m\sqrt{d}}.
\end{equation}
For any $r>p\ge1$ and any matrix $M$, $\|M\|_\infty\le\|M\|_r\le\|M\|_p\le\|M\|_1$ holds, and thus, completes the proof. $\blacksquare$
\bigskip

\bigskip
Furthermore, we need a key lemma known as the McDiarmid's, which is a variant of L\'{e}vy' theorem ({\bf Lemma 3}), defined as follow:

\bigskip
\textbf{Lemma 7} (McDiarmid's inequality~\cite{M89}).
Let $\{X_j\}_{j=1}^m$ be $m$ independent random variables with $X_j$'s chosen uniformly at random from a set $\cl{S}$. Suppose that the measurable function $F:\cl{S}^m\to\bb{R}$ satisfies $|F(x)-F(\hat{x})|\le c_j$, known as the bounded difference, where the vectors $x$ and $\hat{x}$ differ only in the $j$-th position. If we define $Y=F(X_1,\ldots,X_m)$ as a corresponding random variable, then for any $t\ge0$, we have
\begin{equation} \label{McD}
\tn{Pr}[|Y-\bb{E}(Y)|\ge t]\le2\exp\left(-\frac{2t^2}{\sum_{j=1}^mc_j^2}\right),
\end{equation}
where $\tn{Pr}$ denotes the probability.
\bigskip

Here, we consider the bounded difference in Eq.~(23) in \textbf{Lemma~7}. Let the $\varepsilon$-randomizing map $\Lambda$ be realized by a unitary sequence $(U_j)_{j=1}^m$, and another map $\hat{\Lambda}$ be constructed via $(U_1,\ldots,U_{j-1},\hat{U}_j,U_{j+1},\ldots,U_m)$. Thus, we have the difference for the function $F$. That is,
\begin{align*}
\left|\left\|\Lambda(\varphi)-\frac{\1_d}{d}\right\|_p-\left\|\hat{\Lambda}(\varphi)-\frac{\1_d}{d}\right\|_p\right|&\le\left\|\Lambda(\varphi)-\hat{\Lambda}(\varphi)\right\|_p \\
&=\frac{1}{m}\left\|U_j\varphi U_j^\dag-\hat{U}_j\varphi \hat{U}_j^\dag\right\|_p \\
&\le\frac{2^{1/p}}{m}.
\end{align*}

Let us define a random variable $Y_{[\varphi]}:=\left\|\Lambda(\varphi)-\frac{\1_d}{d}\right\|_p$. Thus, the McDiarmid's inequality on the positive part (i.e., $Y_{[\varphi]}-\bb{E}Y_{[\varphi]}>0$) is given by
\begin{equation}
\tn{Pr}\left[Y_{[\varphi]}\ge t+\left(\frac{\sqrt[p]{d}}{m^p}+\frac{r}{m^{p-1}\cdot\sqrt[p]{d}}\right)^{1/r}\right]\le\exp\left(-\frac{mt^2}{2^{(2-p)/p}}\right),
\end{equation}
A similar result can be obtained for the negative part. We can now prove the main proposition.

\subsection{Proof of Theorem 1} \label{app:proofrucp}
Now, we completes the proof of {\bf Thoerem 1}.

\bigskip
{\it Proof}. Let a set $\{U_j\}_{j=1}^m$ be IID $\cl{U}(d)$-valued random variables, distributed according to the unitarily invariant measure. We derive that a map containing $\Lambda(\varphi)=\frac{1}{m}\sum_{j=1}^mU_j\varphi U_j^\dag$ is $\varepsilon$-randomizing with high probability, that is, for any pure quantum state $\varphi\in\cl{B}(\bb{C}^d)$, we find
\begin{equation}
\tn{Pr}_{\forall\varphi}\left[\left\|\frac{1}{m}\sum_{j=1}^mU_j\varphi U_j^\dag-\frac{\1_d}{d}\right\|_p\ge\frac{\varepsilon}{\sqrt[p]{d^{p-1}}}\right]<1.
\end{equation}

If we fix the net $\cl{N}(\bb{C}^d)$ in \textbf{Lemma 2} in the main text and define $\tilde{\varphi}$ to be a net point on the $(d-1)$-sphere corresponding to $\varphi$, then, by the unitary invariance,
\begin{equation}
\left\|\Lambda(\varphi)-\Lambda(\tilde{\varphi})\right\|_1=\left\|\varphi-\tilde{\varphi}\right\|_1\le\frac{\varepsilon}{2\sqrt[p]{d^{p-1}}}.
\end{equation}
In addition, from \textbf{Lemma 2}, we obtain a net with the cardinality $|\cl{N}|\le\left(\frac{10d^{(p-1)/p}}{\varepsilon}\right)^{2d}$. This implies that
\begin{align}
&\tn{Pr}_{\forall\varphi}\left[\left\|\Lambda(\varphi)-\frac{\1_d}{d}\right\|_p\ge\frac{\varepsilon}{d^{(p-1)/p}} \right] \nonumber\\ &\le\tn{Pr}_{\forall\varphi,\tilde{\varphi}}\left[\left\|\Lambda(\varphi)-\Lambda(\tilde{\varphi})\right\|_p+\left\|\Lambda(\tilde{\varphi})-\frac{\1_d}{d}\right\|_p\ge\frac{\varepsilon}{d^{(p-1)/p}}\right] \\
&\le\tn{Pr}_{\forall\tilde{\varphi}}\left[\left\|\Lambda(\tilde{\varphi})-\frac{\1_d}{d}\right\|_p\ge\frac{\varepsilon}{2d^{(p-1)/p}}\right],
\end{align}
because we use $\left\|\Lambda(\varphi)-\Lambda(\tilde{\varphi})\right\|_p\le\left\|\Lambda(\varphi)-\Lambda(\tilde{\varphi})\right\|_1=\left\|\varphi-\tilde{\varphi}\right\|_1\le\frac{\varepsilon}{2\sqrt[p]{d^{p-1}}}$, and the first inequality makes use of the triangle inequality. These inequalities imply that we can change infinitely many pure quantum states to a finite set of pure states (i.e., net points) efficiently.

Finally, by using the union bound, the $\varepsilon$-net (\textbf{Lemma 2}), and the McDiarmid's inequality (\textbf{Lemma 7}) in the main text, we have
\begin{align}
&\tn{Pr}_{\forall\varphi}\left[\left\|\Lambda(\varphi)-\frac{\1_d}{d}\right\|_p\ge\frac{\varepsilon}{d^{(p-1)/p}}\right] \nonumber\\ &\le\tn{Pr}_{\forall\tilde{\varphi}}\left[\left\|\Lambda(\tilde{\varphi})-\frac{\1_d}{d}\right\|_p\ge\frac{\varepsilon}{2d^{(p-1)/p}}\right] \nonumber\\
&\le|N|\cdot\tn{Pr}_{\tilde{\varphi}'}\left[\left\|\Lambda(\tilde{\varphi})-\frac{\1_d}{d}\right\|_p\ge\frac{\varepsilon}{2d^{(p-1)/p}}\right] \\
&\le2\left(\frac{10d^{(p-1)/p}}{\varepsilon}\right)^{2d} \nonumber\\
&\times\exp\left[-\frac{m}{2^{2-2/p}}\left(\frac{\varepsilon}{2d^{(p-1)/p}}-\left(\frac{d^{1/d}}{m^p}+\frac{r}{m^{p-1}d^{1/p}}\right)^{1/r}\right)^2\right]. \nonumber
\end{align}
Thus, there is a $\varepsilon$-randomizing map with respect to the Schatten $p$-norm, if the probability is upper bounded by 1, and $m\ge\frac{\kappa\cdot d}{\varepsilon^2}\log\left(\frac{10d^{(p-1)/p}}{\varepsilon}\right)$. This completes the proof. $\blacksquare$

It is worth noting that for the estimation of the cardinality $m$, we make use of the condition $d<m<d^2$, $r>p\ge1$, and the following probability bound:
\begin{align*}
\left(\frac{10d^{(p-1)/p}}{\varepsilon}\right)^{2d}&\exp\left[-\frac{m}{2^{2-2/p}}\left(\frac{\varepsilon}{2d^{(p-1)/p}}-\frac{2d^{1/rd}}{m^{p/r}}\right)^2\right] \\
&<\frac{1}{2}.
\end{align*}
For sufficiently large $d$, the bound gives rise to $\left(\frac{d^{1/p}}{m^p}+\frac{r}{m^{p-1}d^{1/p}}\right)^{1/r}\le\left(\frac{2d^{1/p}}{m^p}\right)^{1/r}$. Here, if we fix the dimension $d$ and choose $m$ such that $\left(\frac{\varepsilon}{2d^{(p-1)/p}}-\frac{2d^{1/rd}}{m^{p/r}}\right)^2=o(\varepsilon^2)$, then, up to a constant $c$, we have
\begin{equation*}
2d\log\left(\frac{10d^{(p-1)/p}}{\varepsilon}\right)<\frac{cm\varepsilon^2}{2^{2-2/p}}.
\end{equation*}
Therefore, we can conclude that $m\ge\frac{\kappa\cdot d}{\varepsilon^2}\log\frac{10d^{(p-1)/p}}{\varepsilon}$ for a constant $\kappa$.

\subsection{Further components for the reduction of GKZ algorithm} \label{modification}
Here, we briefly introduce two technical lemmas, for the proof of {\bf Proposition 4}, which are the natural modifications of original GKZ algorithm under $\varepsilon$-random technique we suggest. \\

\bigskip
\textbf{Lemma 8} (Modification of Lemma 1~\cite{GKZ19}).
Let $\ket{\mathbf{x}}\in\cl{P}(\bb{C}^d)$ and $\ket{\tilde{\mathbf{x}}}\in\cl{N}(\bb{C}^d)$, respectively. Then we have
\begin{align*}
\tn{\bf TC}~(\mathbf{x}',M)
&=\left\{ \begin{array}{ll} 
1,\ & \;\tn{if}\;\;\mathbf{x}'(=\mathbf{x})=\tilde{\mathbf{x}};\\ & \\
\le(\tfrac{2t+1}{q})^M,\ & \; \tn{if}\;\;\mathbf{x}'(\neq\mathbf{x})\neq\tilde{\mathbf{x}}.
\end{array}\right.
\end{align*}

{\it Proof}. Suppose that there exists a proper quantum state tomography process to read the value $\mathbf{x}$ from the quantum state $\ket{\mathbf{x}}$. By exploiting {\bf Lemma 2} and {\bf Lemma 3} in the main context, we can observe that $\mathbf{x}$ is close to $\tilde{\mathbf{x}}$ with high probability. $\blacksquare$

\bigskip
\textbf{Lemma 9} (Modification of Theorem 1~\cite{GKZ19}).
Let $V\subseteq\bb{F}_q^d$ be a random subset satisfying $|V|=q^{m^*}$ with fixed $m^*=O(\sqrt{d\log d})$ for $m\gg1$.  Let 
\begin{equation*} \label{eq:netqsapp}
|{\tilde{\psi}}\rangle=\frac{1}{\sqrt{|V|}}\sum_{\tilde{\mathbf{a}}\in \bb{F}_q^{m^*}}\ket{\tilde{\mathbf{a}}}\ket{\tilde{\mathbf{a}}\cdot \tilde{\mathbf{x}}+\delta_{\tilde{\mathbf{a}}}\;\tn{mod}~q},
\end{equation*}
where $\delta_{\tilde{\mathbf{a}}}$ is a random variable with absolute value at most $t=poly(m^*)$. Then, the BV algorithm returns $\tilde{\mathbf{x}}$ with probability $\frac{1}{20tq^{m^*-1}}$. \\

%\bigskip
{\it Proof}. By using {\bf Lemma 2} and {\bf Lemma 3} again, it is straightforward, i.e., $\mathbf{a}$ and $\mathbf{x}$ is close to $\tilde{\mathbf{a}}$ and $\tilde{\mathbf{x}}$ with high probability, respectively. $\blacksquare$

%%%%%
%

\end{document}